\documentstyle[12pt]{article}

\def\pd{\partial}
\def\a{\alpha}
\def\b{\beta}
\def\dl{\delta}
\def\s{\sigma}

\def\eps{\epsilon}

\def\lam{\lambda}

\def\bg{{\bar g}}
\def\hg{{\hat g}}

\def\bnabla{{\bar \nabla}}
\def\hnabla{{\hat \nabla}}
\def\nb{\nabla}
\def\bR{{\bar R}}

\def\bF{{\bar F}}
\def\bG{{\bar G}}
\def\bM{{\bar M}}
\def\bL{{\bar L}}
\def\bC{{\bar C}}
\def\bDelta{{\bar \Delta}}
\def\bBox{\stackrel{-}{\Box}}

\def\gm{\gamma}

\def\om{\omega}

\def\sq{\sqrt}
\def\e{\hbox{\large \it e}}
\def\half{\frac{1}{2}}
\def\fr{\frac}
\def\pp{\prime}
\def\arr{\rightarrow}

\def\bb{\begin{equation}}
\def\ee{\end{equation}}
\def\bba{\begin{eqnarray}}
\def\eea{\end{eqnarray}}

\begin{document}

\begin{titlepage}

\begin{tabbing}
   qqqqqqqqqqqqqqqqqqqqqqqqqqqqqqqqqqqqqqqqqqqqqq 
   \= qqqqqqqqqqqqq  \kill 
         \>  {\sc KEK-TH-728}    \\
         \>       hep-th/0012053 \\
         \>  {\sc December, 2000} 
\end{tabbing}
\vspace{1.5cm}

\begin{center}
{\Large {\bf Integrability and Scheme Independence  
of Even-Dimensional Quantum Geometry Effective Action}}
\end{center}

\vspace{1.5cm}

\centering{\sc Ken-ji Hamada\footnote{E-mail address : 
hamada@post.kek.jp} }

\vspace{1cm}

\begin{center}
{\it Institute of Particle and Nuclear Studies, \break 
High Energy Accelerator Research Organization (KEK),} \\ 
{\it Tsukuba, Ibaraki 305-0801, Japan}
\end{center} 

\vspace{1.5cm}

\begin{abstract} 
We investigate how the integrability conditions for  
conformal anomalies constrain the form of the effective action in  
even-dimensional quantum geometry.   
We show that the effective action of four-dimensional quantum 
geometry (4DQG) satisfying integrability has a manifestly 
diffeomorphism invariant and regularization  scheme-independent 
form. We then generalize the arguments 
to six dimensions and propose a model of 6DQG.  
A hypothesized form of the 6DQG effective action is given. 
\end{abstract}
\end{titlepage}  

\section{Introduction}
\setcounter{equation}{0}
\noindent

Since quantum geometry (QG) is defined by functional 
integrations over the metric fields, 
diffeomorphism invariance in QG can equivalently be 
described as an invariance under any change of 
the background metric. 
This background-metric independence includes an invariance 
under a conformal change of the background metric. 
Thus, in even-dimensional QG well-defined on the 
background metric~\cite{p}--\cite{or}, 
conformal anomalies~\cite{cd}--\cite{ii} 
play an important role.  
Therefore, to preserve diffeomorphism invariance 
we must formulate an even-dimensional QG while considering that 
conformal anomalies always exist~\cite{p}--\cite{or}. 

Background-metric independence in two dimensions implies  
that QG can be described as a conformal field theory~\cite{kpz,dk}. 
This idea can be generalized to an arbitrary numbers of 
even dimensions~\cite{amm,hs,h2,kmm,or}. 
However, as recently studied~\cite{hs,h2}, this generalization 
is not simple, because the traceless mode becomes dynamical in higher 
dimensions, so that higher-dimensional QG can no longer be described 
as a free theory. Furthermore, it has been found that 
the integrability condition of the conformal anomaly~\cite{bcr,bpb} 
introduces a strong constraint on even-dimensional QG~\cite{r,ft1,h2}.    

In this paper we further consider how the integrability 
condition of the conformal anomaly affects even-dimensional QG.    
We also settle the problem of the regularization scheme dependence  
and show that the effective action has a manifestly diffeomorphism 
invariant and regularization scheme-independent form. 
 
This paper is organized as follows. In the next section we present  
the fundamental idea of how to preserve diffeomorphism invariance 
in even-dimensional QG and review how such an idea is realized in 
exactly solvable 2DQG~\cite{kpz,dk,s}.  
In $D \geq 4$ dimensions, the integrability condition 
of the conformal anomaly not only restricts interactions of matter 
fields to conformally invariant ones, but also reduces the number 
of indefinite coefficients in the gravity sector~\cite{h2}.  
How the integrability condition affects 4DQG is rediscussed 
in section 3. We then show that the effective action can be written 
in a diffeomorphism invariant and scheme independent form. 
A generalization to six dimensions~\cite{kmm,or} 
is discussed in section 4. 
We show that Duff's scheme~\cite{duff1} 
is also useful to  tame the trivial anomalies 
in six dimensions~\cite{ds}--\cite{bcn}. 
We then propose a model of 6DQG that is  
based on the arguments concerning integrability made in the 4DQG case. 
Many indefinite coefficients that result from  
the existence of many curvature invariants are fixed 
by enforcing the integrability, and a hypothesized scheme-independent 
form of 6DQG effective action is given.  
Section 5 is devoted to conclusions and discussion. 

We use the curvature convention in which 
$R_{\mu\nu}=R^{\lam}_{~\mu\lam\nu}$ 
and $R^{\lam}_{~\mu\s\nu}=\pd_{\s} \Gamma^{\lam}_{~\mu\nu} -\cdots$. 

\section{Conditions of Diffeomorphism Invariance}
\setcounter{equation}{0}
\noindent

In this section we briefly explain how to realize diffeomorphism 
invariance in even-dimensional QG.  

 QG is defined by functional integration over the metric field as 
\bb 
     Z= \int \frac{[g^{-1}d g]_{g} 
                    [dX]_{g} }{\hbox{vol(diff.)}} 
                 \exp \bigl[- I(X,g) \bigr] ~,
                    \label{z} 
\ee
where $I$ is an invariant action and $X$ is a matter field. 
In this paper we consider a conformal scalar without 
self-interactions, for example.  
The measure of the metric field is defined by the invariant 
norm 
\bb
      <dg,dg>_g = \int d^D x \sq{g}g^{\mu\nu}g^{\lam\s} 
           (dg_{\mu\lam}dg_{\nu\s}+udg_{\mu\nu}dg_{\lam\s}) ~, 
\ee
where $D=2n$ and $u >-1/D$. This measure can be orthogonally decomposed 
into the conformal mode and the traceless mode as 
\bba
    && <d \phi, d \phi>_g 
        = \int d^D x \sq{g} (d \phi)^2 ~,
            \label{cmoc}        \\
    && <d h, d h>_g 
        = \int d^D x \sq{g} ~tr (\e^{-h} d \e^h)^2 ~. 
            \label{tmoc}
\eea
Here, the metric is decomposed as $g_{\mu\nu}=\e^{2\phi}\bg_{\mu\nu}$ 
and $\bg_{\mu\nu}=(\hg \e^h)_{\mu\nu}$, 
where $tr(h)=0$~\cite{kkn,hs,h2}.

This definition is manifestly diffeomorphism 
invariant/background-metric independent. However, 
it is not well-defined  because the measures of the metric fields  
defined by (\ref{cmoc}) and (\ref{tmoc}) have a   
metric dependence, represented by $\sq{g}$, in the measures themself,  
so that we must integrate this dependence when we quantize 
the conformal mode $\phi$. 

Instead, we consider measures defined on the background metric as  
\bba
    && <d \phi, d \phi>_{\hg} 
        = \int d^D x \sq{\hg} (d \phi)^2 ~,
           \label{moc}     \\
    && <d h, d h>_{\hg} 
        = \int d^D x \sq{\hg} ~tr (\e^{-h} d \e^h)^2 ~. 
           \label{moh}
\eea
This replacement, however, violates diffeomorphism invariance. 
In fact, these norms conformally change under a general coordinate 
transformation generated by   
$\dl g_{\mu\nu} =g_{\mu\lam}\nabla_{\nu}\xi^{\lam} 
+g_{\nu\lam}\nabla_{\mu}\xi^{\lam}$, 
which can be decomposed as 
\bba
      && \dl \phi = \frac{1}{D} \hnabla_{\lam}\xi^{\lam} +  
                     \xi^{\lam}\pd_{\lam}\phi ~, 
                      \nonumber   \\ 
      && \dl \bg_{\mu\nu} = \bg_{\mu\lam}\bnabla_{\nu}\xi^{\lam} 
                 +\bg_{\nu\lam}\bnabla_{\mu}\xi^{\lam}  
                 -\frac{2}{D}\bg_{\mu\nu}\hnabla_{\lam}\xi^{\lam} ~,    
                      \label{gct}
\eea  
where the relation 
$\bnabla_{\lam} \xi^{\lam}= \hnabla_{\lam} \xi^{\lam}$ is used.
Therefore, these measures produce conformal 
anomalies~\cite{fujikawa} under the general coordinate transformation. 

As a lesson from 2DQG~\cite{dk,s,h1}, 
in order to preserve diffeomorphism invariance, 
we must add an action $S$ as 
\bb
     Z= \int \frac{[d\phi]_{\hg} [\e^{-h}d \e^h]_{\hg} 
                    [dX]_{\hg} }{\hbox{vol(diff.)}} 
                 \exp \bigl[-S(\phi,\bg) - I(X,g) \bigr] ~, 
                    \label{zz}
\ee   
where the measures of the metric fields are now defined by 
(\ref{moc}) and (\ref{moh}).
  
  Let us now briefly see how background-metric independence constrains 
the theory provided by (\ref{zz}). 
Background-metric independence for the traceless mode represents 
the condition that $\hg$ and $h$ always appear in the combination 
$\bg=\hg\e^h$ in (\ref{zz})~\cite{hs}.  
This condition guarantees, at most, that the effective action has  
an invariant form on the metric $\bg$.      
  
  Background-metric independence for the conformal mode 
requires that $S$ satisfies 
the Wess-Zumino condition~\cite{wzc}, 
defined by 
\bb
     S(\phi, \bg) =S(\om, \bg)+S(\phi-\om,\e^{2\om}\bg) ~. 
              \label{wz}
\ee 
Such an action is obtained by integrating the conformal anomaly 
within the interval $[0,\phi]$. Hence it satisfies the initial 
condition $S(0,\bg)=0$ and has a local form. 
In this paper we call this local action the Wess-Zumino action, 
because condition (\ref{wz}) is essential in the arguments  
concerning  diffeomorphism invariance. 
In two dimensions it is usually called 
the Liouville action~\cite{p}.      
The well-known non-local forms of the integrated 
conformal anomaly are called Polyakov action~\cite{p} and 
Riegert action~\cite{r} in two and four dimensions, respectively.  
Why we distinguish between the local and the non-local actions 
becomes clear below. 

Although the Wess-Zumino condition fixes the form of $S$,    
some overall coefficients remain to be determined.  
These coefficients should be determined from the requirement of  
diffeomorphism invariance in a self-consistent manner.   
The process used to determine them is as follows. 

Under the general coordinate transformation, $\dl I=0$, 
while the Wess-Zumino action is not invariant and produces 
a conformal anomaly. 
This property results from condition (\ref{wz}).   
Diffeomorphism invariance is now realized dynamically in such 
a manner that $\dl S$ cancels conformal anomalies  calculated 
with loop effects of the combined theory, ${\cal I}=S+I$.
In other words, we consider the regularized 1PI effective action  
$\Gamma$ of the combined theory ${\cal I}$ 
and require $\dl \Gamma=0$ to determine $S$. 
This means that the tree action ${\cal I}$ is 
not manifestly invariant, but by including loop effects   
the effective action becomes an invariant form on the metric $g$.   

Here, it is worth commenting on the difference in 
the Wess-Zumino action defined by (\ref{wz}) and the non-local 
Polyakov/Riegert action.  
The former produces conformal anomalies 
under a general coordinate transformation, 
while the non-local Polyakov/Riegert action, which appears 
in the effective action due to loop effects, is generally 
defined by the condition that it produces conformal anomalies  
under a conformal change.      
 
  As an exercise, let us first discuss 2DQG coupled $N$ 
conformal scalars.
The tree action in the conformal gauge is given by~\cite{dk,s,h1} 
\bb
    {\cal I} = \fr{b}{4\pi} \int d^2 x \sq{\bg} 
    \bigl( \bg^{\mu\nu}\pd_{\mu} \phi \pd_{\nu} \phi 
     + \bR \phi \bigr) + I_{GF+FP} + I_M(X,\bg) ~,
\ee 
where $I_M$ is the invariant action of the $N$ free scalars.  
The gauge-fixing term and the Faddeev-Popov (FP) ghost action 
are given by~\cite{kato}  
\bb
      I_{GF+FP} 
       = \fr{1}{4\pi} \int d^2 x \sq{\bg} 
    \Bigl( -i B_{\mu\nu}(\bg^{\mu\nu} -\hg^{\mu\nu}) 
     +2 \bg^{\mu\nu} b_{\mu\lam} \bnabla_{\nu} c^{\lam} \Bigr) ~,  
\ee
where the reparametrization ghost $c^{\mu}$ is a contravariant vector.  
$B_{\mu\nu}$ and the anti-ghost $b_{\mu\nu}$ are covariant symmetric 
traceless tensors.  
The coefficient $b$ is uniquely determined by diffeomorphism invariance.  

  Consider the effective action of 2DQG, which has the form
\bb
    \Gamma = {\cal I}(\phi,X,\bg) + W(\bg) ~, 
\ee
where $W$ is a loop effect that depends only on $\bg$ 
because the measure is now  defined on $\bg$. 
The condition of diffeomorphism 
invaraiance, $\dl \Gamma=0$, is now given by 
\bb
         -\fr{b}{4\pi} \int d^2 x \sq{\bg} \om \bR 
          + \dl_{\om} W(\bg) =0 ~,      
\ee 
where $\dl_{\om}\bg_{\mu\nu}=2\om \bg_{\mu\nu}$ and 
$\om=-\half \hnabla_{\lam}{\xi}^{\lam}$. Here, $\dl W = \dl_{\om}W$ 
because $W$ does not depend on the conformal mode $\phi$. 
The second term on the l.h.s is just the conformal anomaly of 
the theory ${\cal I}$. 

{} From one-loop calculations using the tree action ${\cal I}$, we 
obtain the well-known non-local Polyakov action~\cite{p},
\bb
       W(\bg) = \fr{N-25}{96\pi}\int d^2 x 
            \sq{\bg}\bR \fr{1}{\bBox} \bR ~, 
\ee
where $N$ comes from scalar matter fields and this becomes 
$N-26$ through the effect of the ghosts.  
The change in the coefficient from $N-26$ to $N-25$ is due to 
a contribution from the conformal mode. 

As mentioned above, diffeomorphism invariance determines 
the coefficient $b$ uniquely as~\cite{dk}  
\bb
     b=\fr{25-N}{6} ~. 
        \label{b}
\ee 
Using the relation
\bba
   &&  -\fr{1}{24\pi} \int d^2 x \sq{\bg} 
    \bigl( \bg^{\mu\nu}\pd_{\mu} \phi \pd_{\nu} \phi 
     + \bR \phi \bigr) + \fr{1}{96\pi}\int d^2 x 
            \sq{\bg}\bR \fr{1}{\bBox} \bR 
              \nonumber \\ 
   &&  =\fr{1}{96\pi}\int d^2 x 
            \sq{g} R \fr{1}{\Box} R ~,
\eea 
the effective action can be reexpressed in a manifestly invariant 
form, 
\bb
    \Gamma = \fr{N-25}{96\pi}\int d^2 x 
            \sq{g} R \fr{1}{\Box} R + I_M (X,g) ~. 
\ee  
Here, we have used the fact that the matter action is conformally 
invariant, so that $I_M (X,\bg)=I_M (X,g)$.

\section{4D Quantum Geometry}
\setcounter{equation}{0}
\noindent

Recently, we showed that there is a model of diffeomorphism 
invariant 4DQG~\cite{hs,h2}. 
This model has many advantages in physics. In particular, 
it is renormalizable and asymptotically free.   
Also, it is capable of solving the cosmological constant 
problem dynamically 
without any fine-tuning,~\cite{t1,am} and  
it naturally describes our four dimensional universe 
at the low-energy region and for   
large $N$.~\footnote{
In contrast to 2DQG, in which the classical limit is given by 
$N \arr -\infty$, the limit of positive large $N$ gives the correct 
classical limit in 4DQG~\cite{am}.  
} 
However,  the unitarity problem remains unsolved. 
In this paper we do not discuss the unitarity problem, 
which is expected to be solved dynamically~\cite{t2,ft2,kaku,h2}.  

\subsection{Tree action}
\noindent

The tree action of 4DQG~\cite{hs} is given  by a proper combination of 
the Wess-Zumino action~\cite{r,ft1} 
and the invariant action required by the 
integrability conditions discussed in~\cite{h2} and also 
in the following subsection 3.3 as   
\bba
   &&{\cal I}= \frac{1}{(4\pi)^2} \int d^4 x \sq{\bg}
        \biggl\{ ~\frac{1}{t^2}{\bar F} + a {\bar F} \phi 
          + 2 b \phi  \bDelta_4 \phi 
          + b \Bigl( {\bar G}-\fr{2}{3} \bBox {\bar R} \Bigr) \phi 
              \nonumber   \\
   && \qquad\qquad\quad
        + \frac{1}{36}(2a+2b+3c){\bar R}^2 
        + {\cal L}_{GF+FP}  \biggr\} + I_{LE}(X,g) ~,
                \label{tree}
\eea 
where ${\cal L}_{GF+FP}$ contains  the gauge-fixing 
term and the FP ghost Lagrangian defined below. 
The term $I_{LE}$ represents lower derivative actions which  
include actions of conformally invariant matter 
fields, the Einstein-Hilbert action, and the cosmological constant 
term.  The lower derivative gravitational actions  
are treated in the perturbation of the massive constants~\cite{am,ao,h2}.  

The invariants $F$ and $G$ are defined by  
\bba
    F &=& R_{\mu\nu\lam\s}R^{\mu\nu\lam\s}-2R_{\mu\nu}R^{\mu\nu} 
           +\fr{1}{3}R^2 ~,
               \\ 
    G &=& R_{\mu\nu\lam\s}R^{\mu\nu\lam\s}-4R_{\mu\nu}R^{\mu\nu} 
           +R^2 ~. 
\eea 
In four dimensions they are the square of the Weyl tensor and 
the Euler density, respectively. The operator    
$\Delta_4$ is the conformally covariant fourth-order operator~\cite{r},  
\bb
      \Delta_4 = \Box^2 
               + 2 R^{\mu\nu}\nabla_{\mu}\nabla_{\nu} 
                -\fr{2}{3}R \Box 
                + \fr{1}{3}(\nabla^{\mu}R)\nabla_{\mu} ~,  
\ee
which satisfies $\Delta_4 = \e^{-4\phi}\bDelta_4 $ locally for a scalar. 

 In the above, we introduce the dimensionless coupling $t$ 
only for the traceless mode as $\bg_{\mu\nu}=(\hg \e^{th})_{\mu\nu}$ 
and consider the perturbation of $t$. 
The kinetic term of the conformal mode comes from the Wess-Zumino 
action. Since the invariant $R^2$ terms in the Wess-Zumino action 
and the invariant action cancel out in our model, 
the self-interactions of $\phi$ appear only 
in the lower derivative actions in the exponential form, 
which can be treated exactly, order by order in $t$~\cite{h2}.   
 
 The gauge-fixing term and the FP ghost action are  
given by~\cite{ft2,bjs}.  
\bb
      {\cal L}_{GF+FP}  = 2i B^{\mu} N_{\mu\nu} \chi^{\nu}   
            - \zeta B^{\mu} N_{\mu\nu} B^{\nu}    
            -2i {\tilde c}^{\mu} N_{\mu\nu} \hnabla^{\lam} 
              {\bf \dl_B} h^{\nu}_{~\lam} ~, 
                          \label{gfix}
\ee
where $\chi^{\nu}=\hnabla^{\lam}h^{\nu}_{~\lam}$, and $N_{\mu\nu}$ is a 
symmetric second-order operator. 
The BRST transformations are given by 
\bba
    && {\bf \dl_B} h^{\mu}_{~\nu}  
       = i \biggl\{ \hnabla^{\mu} c_{\nu} 
                       +\hnabla_{\nu} c^{\mu}
           - \half \dl^{\mu}_{~\nu}  
                    \hnabla_{\lam} c^{\lam}   
           + t c^{\lam} \hnabla_{\lam} h^{\mu}_{~\nu}   
                \nonumber    \\ 
    && \qquad 
           + \frac{t}{2} h^{\mu}_{~\lam} 
                \Bigl( \hnabla_{\nu} c^{\lam} 
                  - \hnabla^{\lam} c_{\nu} \Bigr) 
           + \frac{t}{2} h^{\lam}_{~\nu} 
                \Bigl( \hnabla^{\mu} c_{\lam} 
                  - \hnabla_{\lam} c^{\mu} \Bigr) 
           + \cdots \biggr\}  ~,
               \nonumber   \\ 
    && {\bf \dl_B} \phi 
       = i t c^{\lam} \pd_{\lam} \phi 
         + i\frac{t}{4} \hnabla_{\lam} c^{\lam} ~, 
            \label{brst} \\ 
    && {\bf \dl_B} {\tilde c}^{\mu} =  B^{\mu} ~,  \qquad
       {\bf \dl_B} B^{\mu} = 0 ~, 
              \nonumber     \\ 
    && {\bf \dl_B} c^{\mu} 
       = i t c^{\lam}\hnabla_{\lam} c^{\mu}  ~.   
            \nonumber
\eea
The first two of these equations are obtained by replacing $\xi^{\mu}/t$ 
in the equation for general coordinate transformation, (\ref{gct}), with 
the contravariant vector ghost field $ic^{\mu}$.  
The kinetic term of the ghost  
action then becomes $t$ independent. 
This BRST transformation is nilpotent. 
Using this transformation, the gauge-fixing term and the FP ghost 
action can be written as ${\cal L}_{GF+FP}= 2i{\bf \dl_B} 
\{ {\tilde c}^{\mu}N_{\mu\nu} (\chi^{\nu} 
+ \frac{i}{2}\zeta B^{\nu}) \}$~\cite{ku}.

The important property of this tree action is that it transforms 
under the general coordinate transformation (\ref{gct}) as 
\bb
   \dl {\cal I} = \fr{1}{(4\pi)^2} \int d^4 x \sq{\bg} 
      ~\om \biggl\{ -a \biggl( \bF + \fr{2}{3}\bBox \bR \biggr) 
                   -b \bG - c \bBox \bR \biggr\} ~, 
              \label{vta}
\ee
where 
\bb
         \om = -\fr{1}{4} \hnabla_{\lam} \xi^{\lam} ~. 
              \label{om4}
\ee 
In the case of the BRST transformation, $\xi^{\mu}$ is replaced 
by $itc^{\mu}$.~\footnote{
Even in 2DQG, although we can set ${\bf \dl_B} {\cal I}=0$ 
if we use the flat background-metric and integrate out the 
$B_{\mu\nu}$ field, 
the nilpotency of the BRST charge at the quantum level, 
after all, requires condition (\ref{b}). 
Thus, the BRST invariance in even-dimensional QG is 
realized dynamically.   
} 

The $\bBox \bR$ terms in (\ref{vta}) depend on the regularization scheme. 
We use here Duff's scheme~\cite{duff1} of dimensional regularization 
characterized by the equations
\bba
   && \dl_{\phi} \int d^D x \sq{g} F 
      = (D-4) \int d^D x \sq{g} \phi \biggl( 
          F +\fr{2}{3} \Box R \biggr) ~, 
          \\ 
   &&  \dl_{\phi} \int d^D x \sq{g} G 
      = (D-4) \int d^D x \sq{g} \phi G ~.  
\eea 
When we define the tree action ${\cal I}$,  
it is taken into account that   
Duff's scheme will be used subsequently for computing loop effects 
of the effective action.  
As shown below, the scheme-dependent terms cancel out, and 
we obtain a scheme-independent effective action.

\subsection{Effective action} 
\noindent

As investigated in~\cite{h2},~\footnote{
Some errors in the form of the effective action in section 3.3 
of ref.~\cite{h2} are corrected in this section. 
} 
the regularized effective action of the 
theory ${\cal I}$ has the following form:  
\bb
   \Gamma = {\cal I}(X,\phi, \bg) 
         +V_{NS}(\phi,\bg) + W_F(\bg,\mu) + W_G(\bg) +W_{\Box R}(\bg)~.
              \label{eff4}
\ee   
Here, the first term on the r.h.s. is the tree action.  $V_{NS}$,  
$W_F$, $W_G$ and $W_{\Box R}$ come from loop diagrams. 
The former represents corrections to the Wess-Zumino action, 
and the latter three represent  
corrections to the traceless mode $h$.   

Let us first consider corrections to the traceless mode. 
Here, $W_F$ is the part that is associated with the conformally invariant 
counterterm of $\bF$; it can be determined by computing two-point 
diagrams of the traceless mode. 
In Duff's scheme, it has the following scale-dependent form: 
\bb
    W_F(\bg,\mu) = \frac{f}{(4\pi)^2}\int d^4 x \sq{\bg} 
           \biggl\{ -\fr{1}{4} \bC_{\mu\nu\lam\s} 
           \log \biggl( \fr{\bDelta^C_4}{\mu^4} \biggr) 
            \bC^{\mu\nu\lam\s} - \frac{1}{18} \bR^2 \biggr\} ~.
\ee 
Here, the appearance of the $\bR^2$ term is 
due to our use of Duff's scheme. 
$C$ is the Weyl tensor, and $\Delta^C_4 = \Box^2 + \cdots$ 
is an appropriate conformally covariant operator for the Weyl tensor. 
Although the explicit form of $\Delta^C_4$ 
is unknown, it is known that there is a function $W_F$ that satisfies 
the equation~\cite{ddi,ds,deser} 
\bb
   \dl W_F(\bg,\mu) =\dl_{\om} W_F(\bg,\mu) 
    = \fr{f}{(4\pi)^2} \int d^4 x \sq{\bg} ~\om 
        \biggl( \bF + \fr{2}{3}\bBox \bR \biggr) ~, 
\ee 
where $\dl_{\om}\bg_{\mu\nu}=2\om \bg_{\mu\nu}$, with (\ref{om4}). 
Thus, $W_F$ produces the type-B anomaly in the classification 
of~\cite{ds}.   

The term $W_G$ in (\ref{eff4}) is the part that is associated with 
the conformally invariant counterterm of $\bG$. 
It is called the non-local Riegert action, 
which produces the type-A anomaly, or the Euler density in the 
classification of~\cite{ds},    
and it has the form  
\bb     
   W_G(\bg) =  \frac{e}{(4\pi)^2}\int d^4 x  
            \sq{\bg} \biggl\{ \fr{1}{8}{\bar {\cal G}} 
             \fr{1}{\bDelta}_4  {\bar {\cal G}} 
               -\frac{1}{18} \bR^2  \biggr\} ~,
\ee
where 
\bb
    {\cal G} = G - \frac{2}{3}\Box R  ~. 
\ee
As stated above, $W_G$ produces the type-A anomaly as 
\bb
    \dl W_G(\bg) = \dl_{\om} W_G (\bg) 
      = \fr{e}{(4\pi)^2} \int d^4 x \sq{\bg} ~\om \bG ~. 
            \label{vwg}
\ee
The $\bR^2$ term is needed to realize equation (\ref{vwg}),  
which guarantees that $W_G$ does not have any contribution     
to two-point diagrams of the traceless mode $h$ in the flat 
background. This is consistent with the direct loop calculations 
of two-point diagrams of $h$.  
Hence, $W_G$ is related to $h^3 $ vetex corrections in the 
flat background. 

The coefficients $f$ and $e$ are scheme independent. 
They can be expanded by the renormalized 
coupling $t_r$ as 
\bb 
   f=f_0 + f_1 t_r^2  + \cdots ~, \qquad   
   e=e_0 + e_1 t_r^2  +\cdots ~.   
\ee
Here, $f_0$ and $e_0$ have already been computed using one-loop 
diagrams as 
\bb
     f_0 = -\fr{N}{120}-\fr{199}{30}+\fr{1}{15} ~, \qquad 
     e_0 = \fr{N}{360} +\fr{87}{20} -\fr{7}{90} ~, 
\ee 
where the first term in each coefficient comes from 
$N$ conformal scalar fields~\cite{duff1}. 
The second and the last terms  
come from the traceless mode~\cite{ft2} 
and the conformal mode~\cite{amm}, respectively.   
The coefficients $f_1$ and $e_1$ are given by functions of 
$a$ and $b$, to which not only two-loop diagrams, but also 
one-loop (but order $t_r^2$) diagrams contribute~\cite{h2}.   

The beta function for the coupling $t_r$ is given by 
$\b=\fr{f}{2} t_r^3$.  
Since $f_0$ is negative, 4DQG is asymptotically free. 
Here, note that, although background-metric independence implies 
an invariance under any confromal change of the background metric, 
the usual $\b$ function does not need to vanish. This is due to  
the fact that there exists a conformal anomaly, or the Wess-Zumino 
action. 
 
The last term in (\ref{eff4}) is a scheme-dependent part, defined by 
\bb 
  W_{\Box R}(\bg) 
   = -\fr{u}{12(4\pi)^2} \int d^4 x \sq{\bg}\bR^2 ~. 
\ee 
It is unknown whether this term is really 
necessary or not.  In any case, the coefficient $u$ is at most 
order $t^2$, so that $u= u_1t_r^2  +\cdots$. 

As computed in~\cite{h2}, 
the correction $V_{NS}$ is scale-independent, and it 
merely changes coefficients 
$a$ and $b$ in the tree action into ${\tilde a}=a(1+v_a)$ and 
${\tilde b}=b(1+v_b)$, 
where $v_a$ and $v_b$ are order $t_r^2$ at the one-loop level. 
The implications of this fact are discussed in the following 
subsection.

Now, the conditions for diffeomorphism invariance are given by 
the following equations~\cite{h2}:  
\bb
       {\tilde a}= f ~, \qquad {\tilde b}= e ~, \qquad c = u ~. 
            \label{bmi4}
\ee 
Since $f_1$ and $e_1$ are functions of $a$ and $b$, while $f_0$ 
and $e_0$ are constants independent of $a$ and $b$, 
we can solve these equations perturbatively, order by order in $t_r$. 
Note that the one-loop coefficients of $v_a$ and $v_b$ are related to 
the order $t_r^2$ coefficients, $f_1$ and $e_1$, of $W_F$ and $W_G$. 
This is reasonable, because the Wess-Zumino action originally comes 
from the measure, and thus is essentially a quantum effect. 
Thus, one-loop contributions given by quantizing the Wess-Zumino 
action are related to two-loop contributions.    

Substituting the solutions of (\ref{bmi4}) into the regularized  
effective action,  
the $\bR^2$ terms cancel out, and we obtain the scheme-independent 
and manifestly invariant effective 
action,~\footnote{
The possibility that an invariant $R^2$ term appears in 
the effective action is not excluded. 
There is a possibility that such a term 
appears in $V_{NS}$ at order $t^4_r$.
}
\bb
    \Gamma  
    = \fr{1}{(4\pi)^2}\int d^4 x \sq{g} \biggl\{  
        - \frac{f}{4}  C_{\mu\nu\lam\s} 
           \log\biggl( \fr{\Delta^C_4}{\mu^4} \biggr) C^{\mu\nu\lam\s}         
        +\frac{e}{8} {\cal G} \fr{1}{\Delta_4} {\cal G} \biggr\}
          + I_{LE}(X,g) ~. 
\ee
Here, the Weyl action $F$ is absorbed into the scale, $\mu$. 

\subsection{Two-loop integrability}
\noindent

Here, we summarize the conditions of diffeomorphism invariance 
discussed in ref.~\cite{h2}.  

The condition that a theory can be made diffeomorphism invariant 
is that in the effective action, there is no action which produces 
a term that does not appear in the variation of 
the tree action $\dl {\cal I}$, (\ref{vta}).  
Namely, diffeomorphism invariance implies that  
the action    
\bb 
   W_{R^2}(\bg,\mu) = \fr{r}{(4\pi)^2}\int d^4 x \sq{\bg} 
              \bR \log  \biggl( 
           \fr{\bDelta_4}{\mu^4} \biggr) \bR 
              \label{wr2}
\ee 
is not allowed, because this action produces $\bR^2$ under a general 
coordinate transformation. 
Further, a scale-dependent action including the conformal 
mode $\phi$, for example   
\bb
   V_S (\phi,\bg,\mu) = \fr{s}{(4\pi)^2}\int d^4 x \sq{\bg} 
             \phi \bDelta_4 \log \biggl( 
             \fr{\bDelta_4}{\mu^4} \biggr) \phi ~, 
                \label{wphi}
\ee 
is not allowed,   
because this action cannot be absorbed into the Wess-Zumino action 
by changing coefficients $a$ and $b$, and it produces a term 
that is not in $\dl {\cal I}$ under a general coordinate 
transformation.  

In general, parts of the effective actions, other than that which    
produces the type-B anomaly,  must be indpendent 
of the scale $\mu$, as $V_{NS}$, $W_G$ and $W_{\Box R}$. 
The vanishing of $r$ and $s$, at least to order $t^2_r$,  
is demonstrated in a previous paper~\cite{h2}. 
We give this demonstration in the following.  
  
First, we expand $r$ and $s$ as $r=r_0 + r_1 t^2_r +\cdots$ 
and $s=s_0 + s_1 t^2_r +\cdots$. 
The vanishing of $r_0$ is guaranteed in our model because 
at this order, only conformally invariant vertices  
contribute to the one-loop diagrams. 
This is a consequence of the fact that the invariant $R^2$ 
terms with the coefficients $a$, $b$ and $c$  
cancel out, so that self-interactions of the conformal 
mode $\phi$ do not appear 
in the tree action ${\cal I}$, except in the 
lower derivative terms, such as the cosmological constant 
in the exponential form. This fact also implies $s_0=0$, 
because there are no diagrams that contribute to $s_0$. 

The vanishing of $s_1$ is proved directly by showing the 
finiteness of  the self-energy diagram of $\phi$~\cite{h2}. 
Here, the fact that there are no interactions of $R^2$ is 
also essential.  
Note that we cannot explain this result by using conformal 
invariance, because conformal invariance does not forbid 
that there exists the counterterm of $\phi \bDelta_4 \phi$. 
It can be explained only by diffeomorphism 
invariance/background-metric independence.   
 
The background-metric independence for the conformal mode 
implies that $W_{R^2}$ and $V_S$ are related in such a manner  
that $s=0$ implies $r=0$.  
Thus, $r_1=0$ is shown indirectly. 

A more direct demonstration of $r_1=0$ is as follows. 
Since there are no self-interactions of $\phi$,
two-loop diagrams that contribute to $f_1$, $g_1$ and $r_1$ 
can be derived from the conformally invariant vertices of  
$2b\phi \bDelta_4 \phi$ and $\fr{1}{t^2}\bF$,  
so that the contributions of two-loop diagrams to $r_1$ vanish. 
However, there are contributions from one loop (but order $t_r^2$)  
diagrams, which include the vertices of $a\bF\phi$, 
$b(\bG-\fr{2}{3}\bBox\bR)\phi$ and $\fr{1}{32}(2a+2b+3c)\bR^2$. 
Here, because these vertices, with the exception of  the first 
one, are non-conformally invariant, we must pay attention to 
such one-loop contributions.     

As shown in~\cite{hs,h1}, the variation in the one-loop contributions 
to the effective action  of our model is given by 
\bb
    \dl_{\om}W^{(1)}(\hg) = -2 {\rm Tr}(\om \e^{-\eps{\cal K}}) ~,
         \label{kernel}
\ee
where $\eps$ is a cutoff. 
The matrix operator ${\cal K}$ is defined by the kinetic term 
$\half \Phi^t {\cal K} \Phi$ on an arbitrary background-metric $\hg$, 
where $\Phi=(\phi, h^{\mu}_{~\nu}, X)$. 
The $t$-independent diagonal parts give the coefficients $f_0$ 
and $e_0$.  The off-diagonal parts, as well as the $t$-dependent diagonal 
parts, give contributions of order $t^2$.   
Note that, unlike in the case of matter fields, we do not use the 
condition of conformal invariance for gravitational fields to derive 
this expression. 
We merely use the facts that ${\cal K}$ is a fourth-order operator and 
there are no self-interactions of the conformal mode. 
If there are the invariant $R^2$ term with the coefficients $a$, $b$  
and $c$, we cannot describe $\dl_{\om}W^{(1)}$ in such a simple form, 
because we do not introduce the coupling for the conformal 
mode $\phi$. That $\dl_{\om}W^{(1)}$ is expressed in the simple form 
such as the r.h.s. of (\ref{kernel}) is a general property of $2n$-th 
order operators in $2n$ dimensions, and such a quantity has been 
shown to be integrable~\cite{hs,h2}.  
Thus, our model satisfies $r_1=0$.    

In four dimensions, integrability places strong constraints on QG. 
It seems that there is no 4DQG other than ours that overcomes  
the integrability conditions.
Thus, 4DQG may be fixed uniquely according to the conformal matter 
content.       

\section{6D Quantum Geometry}
\setcounter{equation}{0}
\noindent

 In this section we show that the arguments concerning integrability 
in 4DQG can be generalized to the six dimensional case. 
Since there are many curvature invariants in six dimensions, 
many indefinite coefficients appear in the definition of 6D action.   
However, we show below that many of them are fixed by the 
integrability.  

\subsection{Duff's scheme in six dimensions}
\noindent

Recently, six-dimensional conformal anomalies have been studied 
in detail~\cite{bpb}--\cite{ii}. 
In this subsection we summarize the results of these studies 
and then show that we can also apply Duff's scheme 
to the six-dimensional case.  

In six dimensions there are 17 independent curvature invariants. 
We here use the following bases~\cite{bpb,bcn}: 
\bba
  && K_1=R^3~, \qquad\quad K_2=RR_{\mu\nu}R^{\mu\nu}~, \qquad\quad 
     K_3= RR_{\mu\nu\lam\s}R^{\mu\nu\lam\s}~, 
           \nonumber  \\ 
  && K_4=R_{\mu}^{~\nu}R_{\nu}^{~\lam}R_{\lam}^{~\mu} ~, \qquad 
     K_5=R_{\mu\nu}R_{\lam\s}R^{\mu\lam\s\nu}~, \qquad 
     K_6= R_{\mu\nu}R^{\mu}_{~\a\b\gm}R^{\nu\a\b\gm}~,          
           \nonumber  \\ 
  && K_7=R_{\mu\nu}^{~~\a\b}R_{\a\b}^{~~\lam\s}R_{\lam\s}^{~~\mu\nu} ~, 
      \qquad 
     K_8=R_{\mu\a\b\nu}R^{\a\lam\s\b}R_{\lam~~\s}^{~\mu\nu}~, \qquad 
     K_9= R\Box R ~,          
           \nonumber \\ 
  && K_{10}=R_{\mu\nu}\Box R^{\mu\nu} ~, \qquad 
     K_{11}=R_{\mu\nu\lam\s}\Box R^{\mu\nu\lam\s} ~, \qquad 
     K_{12}=R^{\mu\nu}\nb_{\mu}\nb_{\nu} R ~, \qquad 
           \nonumber  \\ 
  && K_{13}=(\nb_{\lam}R_{\mu\nu})\nb^{\lam}R^{\mu\nu} ~, \qquad 
     K_{14}=(\nb_{\lam}R_{\mu\nu})\nb^{\mu}R^{\nu\lam} ~,   
               \nonumber \\ 
  && K_{15}=(\nb_{\lam}R_{\a\b\gm\dl})\nb^{\lam}R^{\a\b\gm\dl} ~, \quad 
     K_{16}=\Box R^2 ~, \quad  K_{17}=\Box^2 R ~. 
\eea
 
The results for conformal anomalies are summarized as follows. 
There are ten independent integrable curvature invariants~\cite{bcn}. 
They provide a basis for the conformal anomalies in six dimensions. 
In the classification of ref.~\cite{ds}, the type-A anomaly is 
unique and given by the Euler density,  
\bb 
   G_6 = -K_1 +12 K_2 -3 K_3 -16 K_4 +24 K_5 +24 K_6 
         -4 K_7 -8 K_8 ~. 
\ee 
Here, we normalize it as 
\bb
     G_6 = -\fr{1}{8} \eps_{\mu_1 \nu_1 \mu_2 \nu_2 \mu_3 \nu_3} 
           \eps^{\lam_1 \s_1 \lam_2 \s_2 \lam_3 \s_3} 
           R^{\mu_1 \nu_1}_{~~~~\lam_1 \s_1} 
           R^{\mu_2 \nu_2}_{~~~~\lam_2 \s_2}
           R^{\mu_3 \nu_3}_{~~~~\lam_3 \s_3} ~. 
\ee

There are three type-B anomalies. They are locally conformally invariant 
in six dimensions: 
\bba
    F_1 &=& \fr{19}{800}K_1 -\fr{57}{160}K_2 +\fr{3}{40}K_3 
         +\fr{7}{16}K_4 -\fr{9}{8}K_5 
         -\fr{3}{4}K_6 + K_8 ~,
                  \\ 
    F_2 &=& \fr{9}{200}K_1 -\fr{27}{40}K_2 +\fr{3}{10}K_3 
         +\fr{5}{4}K_4 -\fr{3}{2}K_5 -3 K_6 + K_7 ~,
                   \\ 
    F_3 &=&-\fr{11}{50} K_1 +\fr{27}{10} K_2 -\fr{6}{5} K_3 
           -K_4 +6 K_5 +2 K_7 -8K_8  
               \nonumber \\ 
        && +\fr{3}{5} K_9 -6K_{10} +6 K_{11} +3 K_{13} 
           -6K_{14} +3K_{15} ~. 
\eea
Here, $F_1$ and $F_2$ correspond to two independent combinations of  
the Weyl tensors,   
$C_{\a\mu\nu\b}C^{\mu\lam\s\nu}C_{\lam~~\s}^{~\a\b}$ and 
$C_{\a\b}^{~~\mu\nu}C_{\mu\nu}^{~~\lam\s}C_{\lam\s}^{~~\a\b}$, 
respectively.  
$F_3$ gives the kinetic term of the traceless mode, which is expressed, 
up to a total derivative term, as
$ C_{\mu\a\b\gm} ( \Box \dl^{\mu}_{~\nu} + 4 R^{\mu}_{~\nu} 
- \fr{6}{5}R \dl^{\mu}_{~\nu} ) C^{\nu\a\b\gm}$.  

The other six combinations are given by 
\bba
  && M_5 = 6K_6 -3K_7 +12K_8 +K_{10} -7K_{11} -11K_{13} +12K_{14} 
            -4K_{15} ~, 
                    \\ 
  && M_6 = -\fr{1}{5}K_9 +K_{10} +\fr{2}{5}K_{12} +K_{13} ~, 
                    \\ 
  && M_7 = K_4 +K_5 -\fr{3}{20}K_9 +\fr{4}{5}K_{12} +K_{14} ~, 
                    \\ 
  && M_8 = -\fr{1}{5}K_9 +K_{11} +\fr{2}{5}K_{12} +K_{15} ~, 
                     \\ 
  && M_9 = K_{16} ~, 
                     \\ 
  && M_{10} = K_{17} ~. 
\eea
These are classified as trivial conformal anomalies. 

In order to treat the trivial anomalies $M_5 \cdots M_{10}$, unambiguously, 
we use dimensional regularization. Consider 
the conformal variations of the functions $G_6$, $F_1$, $F_2$ and $F_3$   
defined by the combinations listed above.  
In $D$ dimensions we obtain the equations   
\bb
   \dl_{\phi} \int d^D x \sq{g} G_6 
   = (D-6) \int d^D x \sq{g} \phi G_6 
\ee
and 
\bb
     \dl_{\phi} \int d^D x \sq{g}  F_i  
     = (D-6) \int d^D x \sq{g} \phi \biggl( 
      F_i +\sum_{n=5}^{10}z_{i,n} M_n \biggr) \quad (i=1,2,3) ~,
           \label{duff6}
\ee
where
\bba
  && [z_{1,5}, ~ z_{1,6}, ~z_{1,7}, ~z_{1,8}, ~z_{1,9}, ~z_{1,10}] 
     = [\fr{1}{16}, -\fr{71}{80}, ~ \fr{15}{16}, 
      ~ \fr{13}{40}, ~\fr{159}{3200},~ 0] ~,
                    \\ 
  && [z_{2,5}, ~ z_{2,6}, ~z_{2,7}, ~z_{2,8}, ~z_{2,9}, ~z_{2,10}] 
     = [-\fr{1}{4}, -\fr{1}{20},  -\fr{3}{4}, 
        -\fr{7}{10},  -\fr{51}{800},~ 0] ~, 
                    \\ 
  && [z_{3,5}, ~ z_{3,6}, ~z_{3,7}, ~z_{3,8}, ~z_{3,9}, ~z_{3,10}] 
     = [1 ,~ \fr{1}{5}, ~ 3, ~ \fr{14}{5}, ~\fr{39}{200},~
        \fr{3}{5}] ~.
\eea 
Here, note that the r.h.s. of equation (\ref{duff6}) is expanded 
in terms of $F_i$ itself and the trivial conformal anomalies. 
This equation suggests that Duff's scheme also works well  
in six dimensions. 

\subsection{Tree action}
\noindent

Let us first look for a conformally covariant sixth-order operator 
in six dimensions~\cite{kmm}.  
It can be expanded in terms of the 21 independent operators,  
apart from the $\Box^3$ term, as 
\bba
  &&\Delta_6 = \Box^3 
     + v_1 R^{\mu\nu}\nb_{\mu}\nb_{\nu}\Box 
     +v_2 R \Box^2 
     + v_3 (\nb^{\lam}R^{\mu\nu})\nb_{\lam}\nb_{\mu}\nb_{\nu} 
           \nonumber  \\
  && \quad 
     +v_4 (\nb^{\lam}R)\nb_{\lam}\Box 
     +v_5 (\nb^{\mu}\nb^{\nu}R)\nb_{\mu}\nb_{\nu} 
     +v_6 (\Box R^{\mu\nu})\nb_{\mu}\nb_{\nu} 
          \nonumber   \\ 
  && \quad 
     +v_7 (\Box R)\Box 
     +v_8 R^{\mu}_{~\a\b\gm}R^{\nu\a\b\gm}\nb_{\mu}\nb_{\nu} 
     +v_9 R_{\mu\nu\lam\s}R^{\mu\nu\lam\s}\Box 
          \nonumber \\ 
  && \quad         
     +v_{10} R^{\a\b}R^{\mu~~~\nu}_{~\a\b}\nb_{\mu}\nb_{\nu} 
     +v_{11} R^{\mu\lam}R^{\nu}_{~\lam} \nb_{\mu}\nb_{\nu}  
     +v_{12} R^{\mu\nu}R_{\mu\nu}\Box 
            \nonumber  \\ 
  && \quad
     +v_{13} RR^{\mu\nu}\nb_{\mu}\nb_{\nu} 
     +v_{14} R^2 \Box 
     +v_{15} (\nb^{\lam}\Box R)\nb_{\lam} 
     +v_{16} R_{\a\b\gm\mu}(\nb^{\mu}R^{\a\b\gm\nu})\nb_{\nu} 
            \nonumber  \\ 
  && \quad 
     +v_{17} R^{\mu\nu\lam\s}(\nb_{\mu}R_{\nu\lam})\nb_{\s} 
     +v_{18} R_{\mu\nu}(\nb^{\mu}R^{\nu\lam})\nb_{\lam} 
     +v_{19} R_{\mu\nu}(\nb^{\lam}R^{\mu\nu})\nb_{\lam} 
            \nonumber  \\ 
  && \quad          
     +v_{20} R^{\mu\nu}(\nb_{\mu}R)\nb_{\nu} 
     +v_{21} R(\nb^{\lam}R)\nb_{\lam} ~,  
\eea
{}From the requirement that $\dl_{\phi}(\sq{g}\Delta_6 Y)= 0$ is 
satisfied locally for a scalar $Y$, the coefficients are 
determined as follows: 
\bba
    && v_1 = 4 ~, \quad v_2 =-1 ~, \quad v_3 = 4 ~, \quad 
       v_4 = 0 ~, \quad v_5 = 0 ~, \quad v_6 = 4 ~, \quad 
         \nonumber  \\ 
    && v_7 = -\fr{3}{5}~, \quad v_8 = \zeta_1 ~, \quad  
       v_9 = \zeta_2 ~, \quad 
       v_{10}= \zeta_1 ~, \quad v_{11}= 6-\fr{3}{4}\zeta_1 ~, \quad 
         \nonumber  \\ 
    && v_{12}= -1 +\fr{1}{8}\zeta_1 -\zeta_2 ~, \quad 
       v_{13}= -2 +\fr{1}{4}\zeta_1 ~, \quad 
       v_{14}= \fr{9}{25} -\fr{1}{40}\zeta_1 +\fr{1}{10}\zeta_2 ~, \quad 
         \nonumber  \\ 
    && v_{15}= \fr{2}{5} ~, \quad
       v_{16}= \zeta_1 +4\zeta_2 ~, \quad
       v_{17}= -\zeta_1 ~, \quad 
       v_{18}= 6 +\fr{1}{4}\zeta_1 ~,  
          \nonumber  \\ 
    && v_{19}= -2 -\fr{3}{4}\zeta_1 -2\zeta_2 ~, ~ 
       v_{20}= 1-\fr{1}{8}\zeta_1 ~, ~ 
       v_{21}= -\fr{7}{25} +\fr{3}{40}\zeta_1 +\fr{1}{5}\zeta_2 ~. 
\eea
In six dimensions, $\Delta_6$ is not unique, as the  
two constants $\zeta_1$ and $\zeta_2$ are not determined by  
the conformal property alone. The terms with these arbitrary constants 
are collected, using the Weyl tensor, in the forms  
$\zeta_1 \nb^{\mu}(C_{\mu\a\b\gm}C^{\nu\a\b\gm}\nb_{\nu})$ and 
$\zeta_2 \nb^{\lam}(C_{\a\b\gm\dl}C^{\a\b\gm\dl}\nb_{\lam})$, 
respectively~\cite{kmm}. 

Next, we look for a combination of $G_6$ and $M_n$ that satisfies 
the following conformal property locally: 
\bb 
    \dl_{\phi} \biggl\{ 
    \sq{g} \biggl( G_6 -\sum_{n=5}^{10} w_n M_n \biggr) 
    \biggr\} = 6 \sq{g}  \Delta_6 \phi  ~. 
          \label{cpog}
\ee
This equation determines the coefficients $w_n$ uniquely 
for each $\Delta_6$ with $\zeta_1$ and $\zeta_2$ as 
\bba 
   && w_5 = 1+\fr{1}{4}\zeta_1 ~, \quad 
      w_6 =11+\half\zeta_1-3\zeta_2 ~, \quad
      w_7 = -6 -\fr{3}{4}\zeta_1 ~, 
          \nonumber  \\  
   && w_8 = 1+\zeta_1 +3\zeta_2 ~, \quad
      w_9 = -\fr{21}{100}+\fr{9}{160}\zeta_1 
            +\fr{3}{20}\zeta_2 ~, \quad
      w_{10} = \fr{3}{5} ~. 
\eea

Using equation (\ref{cpog}), 
the Wess-Zumino action defined by integrating the conformal anomalies 
within the interval $[0,\phi]$ are expressed in the form   
\bba
   &&S(\phi,\bg) 
             \nonumber \\
   &&= \fr{1}{(4\pi)^3} \int d^6 x \int^{\phi}_{0} d\phi 
       \sq{g} \biggl\{ \sum^{3}_{i=1}
        a_i \biggl( F_i +\sum_{n=5}^{10}z_{i,n} M_n  \biggr)  
           + b G_6 + \sum_{10}^{5}c_n M_n \biggr\} 
               \nonumber \\ 
    &&= \fr{1}{(4\pi)^3} \int d^6 x \sq{\bg} \biggl\{  
             \sum^{3}_{i=1}a_i \bF_i ~\phi 
             + 3b \phi \bDelta_6 \phi 
             +b \biggl(\bG_6 -\sum_{n=5}^{10} w_n \bM_n \biggr)\phi 
             \biggr\}  
               \nonumber \\ 
      && \qquad\qquad 
            + \sum^{10}_{n=5} 
             \fr{ \sum_{i=1}^{3}a_i z_{i,n} +b w_n +c_n}{(4\pi)^3} 
            \int d^6 x \Bigl( \sq{g} L_n - \sq{\bg} \bL_n \Bigr) ~.  
               \nonumber
\eea
Here, the $L_n$ are local functions given by integrating the $M_n$  
as 
\bb
       \dl_{\phi} \int d^6 x \sq{g} L_n 
       = \int d^6 x \sq{g} \phi M_n 
\ee
such that  
\bba
  && L_5 = \fr{1}{30}K_1 -\fr{1}{4}K_2 +K_6 ~, \qquad 
     L_6 = \fr{1}{100}K_1 -\fr{1}{20}K_2 ~, \qquad 
            \nonumber \\ 
  && L_7 = \fr{37}{6000}K_1 -\fr{7}{150}K_2 +\fr{1}{75}K_3 
           -\fr{1}{10}K_5 -\fr{1}{15}K_6 ~,  
                 \\ 
  && L_8 = \fr{1}{150}K_1 -\fr{1}{20}K_3 ~, \quad 
     L_9 = -\fr{1}{30}K_1 ~, \quad
     L_{10} = \fr{1}{300}K_1 -\fr{1}{20}K_9 ~. 
             \nonumber 
\eea

As discussed in the case of 4DQG, the integrability condition 
suggests that the sixth-order parts of the invariant action $I$ 
should be chosen such that the invariant $L_n$ terms cancel out 
in the sum ${\cal I}=S+I$. Hence, we obtain a 6DQG tree action 
analogous to that in 4DQG as
\bba
  {\cal I} 
   &=& \fr{1}{(4\pi)^3} \int d^6 x \sq{\bg} \biggl\{  
          -\fr{1}{t^2} \Bigl( \bF_3 +\a_1 \bF_1 +\a_2 \bF_2 \Bigr) 
            + \sum^{3}_{i=1}a_i \bF_i ~\phi 
                 \nonumber \\ 
   && \qquad\qquad\qquad 
         + 3b \phi \bDelta_6 \phi 
         +b \biggl(\bG_6 -\sum_{n=5}^{10} w_n \bM_n \biggr)\phi   
                \\ 
   && \qquad\qquad 
            - \sum^{10}_{n=5} \biggl(
            \sum_{i=1}^{3}a_i z_{i,n} +b w_n +c_n \biggr) \bL_n 
            \biggr\} + I_{LE}(X,g)~.  
               \nonumber
\eea
Here, we introduce the dimensionless coupling $t$, as in 4DQG.  
In six dimensions, two extra dimensionless constants  
$\a_1$ and $\a_2$, in addition to $\zeta_1$ and $\zeta_2$ in $\Delta_6$ 
and $w_n$, appear. These constants are not fixed by the arguments of 
the integrability. The constants $t$, $\a_1$ and $\a_2$ are  
renormalized, but $\zeta_1$ and $\zeta_2$ may not be. 

Under a general coordinate transformation, 
this action changes according to   
\bb
   \dl {\cal I} 
   = \fr{1}{(4\pi)^3} \int d^6 x  
       \sq{\bg} ~\om \biggl\{ -\sum^{3}_{i=1}
        a_i \biggl( \bF_i +\sum_{n=5}^{10}z_{i,n} \bM_n  \biggr)  
           - b \bG_6 - \sum_{10}^{5}c_n \bM_n \biggr\} ~, 
\ee
where 
\bb
    \om=-\fr{1}{6}\hnabla_{\lam}\xi^{\lam} ~. 
           \label{om6}
\ee 
 
\subsection{Effective action}
\noindent

It is expected that the effective action of this model has the 
form 
\bb
    \Gamma = {\tilde {\cal I}}(X,\phi,\bg) + W_{G_6}(\bg)  
              + \sum^{3}_{i=1}W_{F_i}(\bg,\mu)  
               + \sum^{10}_{n=5} W_{M_n}(\bg) ~,  
\ee 
where the tilde on ${\cal I}$ denotes the inclusion of finite 
corrections to the Wess-Zumino action described by $V_{NS}$ 
in the four-dimensional model.  
Here, $W_{G_6}$ is the generalization of the non-local 
and scale-independent Polyakov-Riegert action~\cite{a,deser}. 
We find its complete form as  
\bb
    W_{G_6} (\bg) 
      = \fr{e}{(4\pi)^3} \int d^6 x \sq{\bg} \biggl\{ 
         \fr{1}{12} {\bar {\cal G}}_6 \fr{1}{\bDelta_6} 
          {\bar {\cal G}}_6
             +  \sum_{n=5}^{10} w_n \bL_n \biggr\} ~, 
\ee
where
\bb
         {\cal G}_6 = G_6 -\sum_{n=5}^{10} w_n M_n ~. 
\ee
This produces the type-A anomaly, 
\bb
    \dl W_{G_6}(\bg) = \dl_{\om} W_{G_6} (\bg)
      = \fr{e}{(4\pi)^3} \int d^6 x \sq{\bg}~\om \bG_6 ~,                    
\ee
where $\dl_{\om}\bg_{\mu\nu}=2\om\bg_{\mu\nu}$, with (\ref{om6}).
This equation is realized for arbitrary values of $\zeta_1$ 
and $\zeta_2$. These constants, as well as $e$,  
are determined according to matter content. 

 The action $W_{F_i}$, which produces the type-B 
anomaly in Duff's scheme, is defined by 
\bb
    W_{F_i}(\bg,\mu) =f_i \biggl( W^{\pp}_{F_i}(\bg,\mu) 
                + \sum_{n=5}^{10} \fr{z_{i,n}}{(4\pi)^3} 
                    \int d^6 x \sq{\bg} \bL_n \biggr) ~. 
\ee
The $\bL_n$ terms appear in Duff's scheme. 
$W^{\pp}_{F_i}$ is a scale-dependent part defined 
through the equation 
\bb 
     \dl_{\om} W^{\pp}_{F_i}(\bg,\mu) 
        = \fr{1}{(4\pi)^3} \int d^6 x \sq{\bg} ~\om \bF_i ~.                  
\ee
It is known that the coefficients $e$ and $f_i$ are independent 
of the regularization scheme. 

The remaining action, $W_{M_n}$, is a scheme-dependent part defined by 
\bb
     W_{M_n}(\bg) 
       = \fr{u_n}{(4\pi)^3} \int d^6 x \sq{\bg} \bL_n ~.
\ee 
This action produces a trivial anomaly, $\bM_n$. 
As in 4DQG, it is unknown whether this action is really necessary 
or not.  Since the vertices of the tree action at zeroth order in $t$ 
is conformally invariant, the coefficients $u_n$ will be at most 
order $t^2$.  

The conditions for diffeomorphism invariance are now given by
\bb
      {\tilde a_i} =f_i ~, \qquad  {\tilde b} =e ~, 
      \qquad c_n=u_n ~, 
\ee
where the tildes on $a_i$ and $b$ indicate the inclusions 
of corrections to the Wess-Zumino action. As in 4DQG,  
the scheme-dependent terms, $\bL_n$, cancel out, and 
the final expression takes the invariant and scheme-independent 
form 
\bb
    \Gamma = \fr{e}{(4\pi)^3} \int d^6 x \sq{g} 
        ~ \fr{1}{12} {\cal G}_6 \fr{1}{\Delta_6} {\cal G}_6 
    +\sum_{i=1}^{3} f_i W^{\pp}_{F_i}(g,\mu)+ I_{LE}(X,g) ~.  
\ee
The matter contributions to the coefficients $e$ and $f_i$ are  
computed in refs.~\cite{bft,g,ii}.

\section{Conclusions and Discussion}
\noindent

In this paper we have discussed how the integrability conditions for  
conformal anomalies constrain the form of the effective action of 
even-dimensional QG.   
We showed that the effective action of 4DQG satisfying 
such integrability conditions has a manifestly diffeomorphism 
invariant and  scheme-independent form. 
We then generalized the arguments 
to six dimensions and proposed a model for 6DQG. 
The expected scheme-independent form of the effective action 
was presented. 

Now, the role of conformal anomalies in even-dimensional QG is 
naturally understood in terms of background-metric 
independence/diffeomorphism invariance.  
In $D=2n ~(\geq 4)$ dimensions, unlike the case for 2DQG,  
there is no critical matter content  
where the Wess-Zumino action vanishes.  Thus,  
$2n$-dimensional QG is to be necessarily 
$2n$-th order, because of diffeomorphism invariance.  

Background-metric independence does not require the vanishing 
of the usual beta functions in $D\geq 4$ dimensions,  
though it implies invariance under any 
conformal change of the background metric. 
This is due to the fact that there exist conformal anomalies, 
or the Wess-Zumino action, in even dimensions. 
We believe that conformal invariance in physics should be 
re-interpreted in terms of diffeomorphism invariance.  
In this case, the problem of dependence on the regularization 
scheme would disappear. 

In odd dimensions, because there is no conformal anomaly,  
background-metric independence seems to require the theory 
to be finite.  
In three dimensions the Einstein-Hilbert$+$cosmological constant action 
is written in the Chern-Simons action, and its quantum theory  
is expected to be topological~\cite{witten}. 
However, for $D \geq 5$, it is unknown whether odd-dimensional 
QG exists or not.  
Since in odd dimensions, we cannot introduce a dimensionless coupling 
constant, it seems necessary to make the theory  
super-renormalizable.  

There is another approach to QG based on dynamical 
triangulation in two~\cite{wei,bk,adj} and 
four dimensions~\cite{migdal,bbkptt,efhty,adj}. 
It is expected that our model is obtained in the continuum limit 
of such a simplicial QG. 
In this paper we do not discuss quantum corrections of 
the lower-derivative grvitational actions. 
The anomalous dimensions of the gravitational constant and 
the cosmological constant are needed to compare 
the two methods~\cite{am,h2}. 
A project involving detailed comparison in 4DQG between them has 
started~\cite{efhty}. 

Finally, we comment on dimensional regularization. 
Because dimensional regularization violates conformal invariance 
in general, it is not a suitable regularization for a theory in which  
conformal invariance plays an important role. 
Nevertheless, dimensional regularization is still useful, 
because this violation is quite small and it is expected to  
give correct results for sufficiently higher order loops~\cite{hath}. 

There is an assertion that, when using dimensional 
regularization, we can regularize QG defined by (\ref{z}) 
in a manifestly diffeomorphism invariant way if we take great care 
concerning the conformal mode dependence~\cite{kkn}. 
At present, the relation between this approach and ours is unknown.   
Detailed analyses of this relation are important to prove 
renormalizability to all orders. 

The beautiful relations obtained among integrable curvature 
invariants in $D$ dimensions seem to suggest the validity of 
dimensional regularization. 
Our model, at least up to order $t_r^2$,  
gives correct results because of the finiteness of 
the self-energy diagrams of $\phi$, which implies that our 
model is rather insensitive to the conformal mode dependence. 
Whether or not the derived effective action at higher order is 
acceptable will be decided by the condition that 
it possesses a scheme-independent form and does not contain terms 
that violate diffeomorphism invariance,  
such as (\ref{wr2}) and (\ref{wphi}).

\vspace{5mm}

\begin{flushleft}
{\bf Acknowledgements}
\end{flushleft}

  This work was supported in part by the Grant-in-Aid for 
Scientific Research from the Ministry of 
Education, Science and Culture of Japan.

\end{document}